\documentclass{article}
\usepackage[utf8]{inputenc}
\usepackage{natbib,amsmath,amssymb,amsthm}
\usepackage{algorithm,listings}
\usepackage{graphicx}
\usepackage{url}
\usepackage[noend]{algpseudocode}
\usepackage{caption}
\usepackage{tikz,lipsum}
\usepackage{authblk}
\usepackage{appendix} 

\newcommand*\circled[1]{\tikz[baseline=(char.base)]{
            \node[shape=circle,draw,inner sep=2pt] (char) {#1};}}
        
\title{OCEAN: Flexible Feature Set Aggregation for Analysis of Multi-omics Data}

\author[1]{Mitra Ebarahimpoor \footnote{Corresponding Author: m.ebrahimpoor@lumc.nl}}
\author[2]{Ren\'ee  Menezes}
\author[1]{Ningning Xu}
\author[1]{Jelle J. Goeman}

\affil[1]{Department of Biomedical Data Sciences, Leiden University Medical Center, Leiden, The Netherlands}
\affil[2]{Biostatistics Centre, Department of Psychosocial Research and Epidemiology, Netherlands Cancer Institute, Amsterdam, The Netherlands}

\date{}

\begin{document}

\maketitle

\begin{abstract}

Integrated analysis of multi-omics datasets holds great promise for uncovering complex biological processes. However, the large dimension of omics data poses significant interpretability and multiple testing challenges. Simultaneous Enrichment Analysis (SEA) was introduced to address these issues in single-omics analysis, providing an in-built multiple testing correction and enabling simultaneous feature set testing. In this paper, we introduce OCEAN, an extension of SEA to multi-omics data. OCEAN is a flexible approach to analyze potentially all possible two-way feature sets from any pair of genomics datasets. We also propose two new error rates which are in line with the two-way structure of the data and facilitate interpretation of the results. The power and utility of OCEAN is demonstrated by analyzing copy number and gene expression data for breast and colon cancer.

\end{abstract}

\section{Introduction}\label{intro}

Joint analysis of multiple genomics datasets has the potential to offer a deeper understanding of complex biological phenomena and disease processes \citep{Zhang2017,Huang2011}. By combining information from multiple genomic levels such as gene expression, copy number variation and DNA methylation, say, such analyses can unveil the interaction between various molecular layers and identify synergistic relationships \citep{Stranger2005,Bhattacharya2020,Behring2018,Menezes2016,Huang2011}.

There are several methods for joint analysis of omics data, such as regression, correlation, co-expression, Bayesian networks, and machine learning algorithms such as random forest or support vector machines \citep[see][for an overview]{subramanian_2020}. Studying pairwise associations between omics features remains one of the most commonly adopted methods \citep{Gu2008, Kotliarov2009,Huang2011,Lahti2012}. However, the scale of the analysis is an issue: the huge matrix of pairwise associations can be hard to interpret and leads to a massive multiple testing problem. 

When analyzing a single-omics type, an established solution to the scale problem of the data is to consider sets of features, rather than individual ones. This alleviates the multiple testing problem and improves power \citep{Menezes2009, Menezes2016, Khatri2012, Huang2011, Maleki2020, Zhao2023}, and also leverages the existing biological knowledge, providing context and interpretability. Feature sets can be defined based on shared biological characteristics or attributes such as location, function, biological pathway or disease association. Popular feature set resources include Reactome, KEGG, MSigDB, OncoKB, dbSNP and TCGA. The concept of feature sets has been extended to multi-omics analyses. For example, \citet{Chaturvedi2016} proposed a test for association of two-way feature sets: combinations of a feature set in one omics type with a feature set in another, for example for identifying genomic regions that are associated with changes in gene expression of a molecular pathway. 

Simultaneous Enrichment Analysis (SEA) was recently introduced as a novel way of analyzing feature sets in single-omics \citep{Ebrahimpoor2019}. It has several advantages over classical single-omics feature set methods. It has an in-built multiple testing correction, in contrast with classical feature set testing methods which require an additional multiple testing step after the initial analysis. This multiple testing correction automatically includes all possible feature sets, obviating the need to pre-specify the collection of feature sets of interest. This gives great flexibility to researchers: they may decide upon feature sets of interest after seeing the data, while still maintaining proper Type I error control. SEA was shown to have at least comparable power to classical feature set methods when all of the feature sets in Gene Ontology were tested. Furthermore, rather than merely returning a $p$-value for the sets, it quantifies the signal within feature sets. It does this by giving a simultaneous 95\%-confidence lower bound of the true discovery proportion (TDP), i.e., the proportion of truly active features in the set. 


In this paper we explore the extension of SEA to multi-omics. A direct application in terms of two-way feature sets is relatively straightforward and already showcases some of the advantages of SEA. SEA's flexibility of choosing feature sets {\sl after} seeing the data is especially useful in multi-omics, because there is a quadratic number of two-way feature sets to explore. The large number of potential feature set combinations is also expected to make SEA competitive in terms of power. However, the TDP that SEA returns will be in terms of pairs and ignores the matrix structure of the two-way feature sets. The pair-based TDP is not easy to interpret in terms of each of the individual omics features \citep{Luijk2014}. To solve this, we propose alternative definitions of TDP, quantifying the signal in a way that respects the two-way structure. These per-omic TDPs estimate the number of features within a feature set of omic A that are associated with at least one feature in a feature set of omic B (and vice-versa). We show here that SEA can give 95\%-confidence bounds for the two per-omic TDPs, simultaneous with the pair-TDP, without the need for any additional multiple testing correction.


We will first revisit SEA and introduce its direct extension to multi-omics. Subsequently, we present the two novel TDP measures and the way they can be controlled. This control involves the development of a novel branch-and-bound algorithm. We call our novel method  OCEAN, as it is an extension of SEA. We demonstrate the utility of OCEAN with an integrated analysis of copy number (CN) and gene expression (GE) datasets for breast cancer and colon cancer, and show its ability to identify previously known and novel feature sets of interest in cancer research.

\section{Notation}

Though our method is more general, we will restrict our presentation to two omics types for the sake of simplicity. Assume that we have two omics from a cohort of $n$ individuals: omic $A$ and omic $B$, which include $p$ and $q$ features, respectively. $P$ is a $p \times q$ association matrix where the element $p_{jk}$ represents the corresponding $p$-value for the pairwise association, each testing the null hypothesis $H_{0,jk}$: $j$th feature of omic $A$ is not associated with the $k$th feature of omic $B$.  

We are not primarily interested in testing the individual or elementary hypotheses $H_{0,jk}$. Instead, we are interested in testing the null hypothesis for sets of features. Let $S_A \subseteq \{1,\ldots, p\}$ and $S_B \subseteq \{1,\ldots, q\}$ be feature sets within omic $A$ and $B$, respectively. Let $M = \{1,\ldots, p\} \times \{1,\ldots, q\}$, with $m=pq$ elements, and define $S = S_A \times S_B$ as the subset of $M$ of interest. Considering all possible choices of $S_A$ and $S_B$, there are roughly $2^{p+q}$ possible sub-matrices of $P$, or two-way feature sets, that may be of interest.

\section{SEA for multi-omics}\label{sea}

We will review SEA, which was developed for analysis of single-omics \citep{Ebrahimpoor2019}, and present its first extension to multi-omics. SEA contrasts with other feature set testing methods in two important respects. In the first place, it does not test only a limited database of feature sets, but all possible feature sets, and corrects for multiple testing for all of them. This implies that the validity of the method is not compromised if the user peeks at the data before choosing feature sets of interest. Secondly, it does not primarily report a $p$-value, but a much more informative lower $(1-\alpha)$- confidence bound on the true discovery proportion (TDP), i.e., the proportion of true effects present in the feature set. 

SEA is easily generalized to multi-omics, and we will explain it directly in those terms. In this generalization, the TDP is defined as follows. Let $D \subseteq M$ be the unknown set of true discoveries, i.e., the collection of pairs for which there is some association between omic A and omic B. Then the true discovery proportion ($TDP$) for $S$ is defined as 
\[TDP(S)=\frac{d(S)}{|S|},\]
where $d(S)=| D \cap S|$ is number of correctly identified associations (``true discoveries'') within $S$, and $|S|$ is the number of elements in $S$. 

The goal of SEA is to give a lower bound for TDP($S$) for every $S$ of interest while controlling the Type I error. \citet{Goeman2019} derived such simultaneous lower bounds $\bar{d}(S)$ for all $d(S)$ as
\begin{equation}
\label{bound}
\bar{d}(S)= \max_{1\leq u \leq |S|} 1-u+|\{i\in S\colon hp_i \leq u\alpha\}|,
\end{equation}
where
\begin{equation}\label{functionh}
h=\max\{r\in\{0,...,m\}\colon rp_{(m-r+j)}>j\alpha, \text{for }j=1, ..., r\},
\end{equation}
where $p_{(m-r+j)}$ is the $(m-r+j)$th smallest $p$-value with index in $M$.

For this bound, it holds that
\begin{equation}
\label{simbounds}
\mathrm{P}(\bar{d}(S)\leq d(S) \textrm{\ for all $S$})\geq 1-\alpha,
\end{equation}

i.e., $\bar{d}(S)$ is a lower $(1-\alpha)$- confidence bound for the number of true discoveries in $S$, and it immediately follows that $\bar{d}(S)/|S|$ is a $(1-\alpha)$-confidence bound for TDP($S$). Importantly, the bound is simultaneous for all $S$, implying that it remains valid if $S$ is chosen after seeing the data.

Bounds of the form (\ref{simbounds}) can be derived from any Closed Testing procedure \citep{Marcus1976, genovese2006exceedance, Goeman2011}. For the particular bound (\ref{bound}), SEA uses a particular Closed Testing procedure first proposed by \citet{Hommel1988} that is based on a test of \citet{Simes1986}, and extended to TDP by \citet{Goeman2019}. This procedure assumes that the $p$-values are generally positively correlated, an assumption that the FDR procedure of \cite{Benjamini1995} also makes. \citet{Goeman2019} showed that for this procedure $\bar{d}(S)$ has good power for large numbers of hypotheses, and can be efficiently calculated. 
The extension of SEA to multi-omics is very similar to the SEA for a single omic. It allows a researcher to browse through many feature sets of interest in both omics, choosing the most interesting combination in terms of both biology and TDP. This TDP is defined in terms of pairs: it is a lower bound to the proportion of omic $\times$ omic pairs for which an association is truly present. We will refer to the TDP derived in this section as the pair-TDP.

\section{OCEAN}

\begin{figure}[ht]
	\captionsetup{width=.8\linewidth}
    \centering
	\includegraphics[width=0.8\linewidth,keepaspectratio=true]{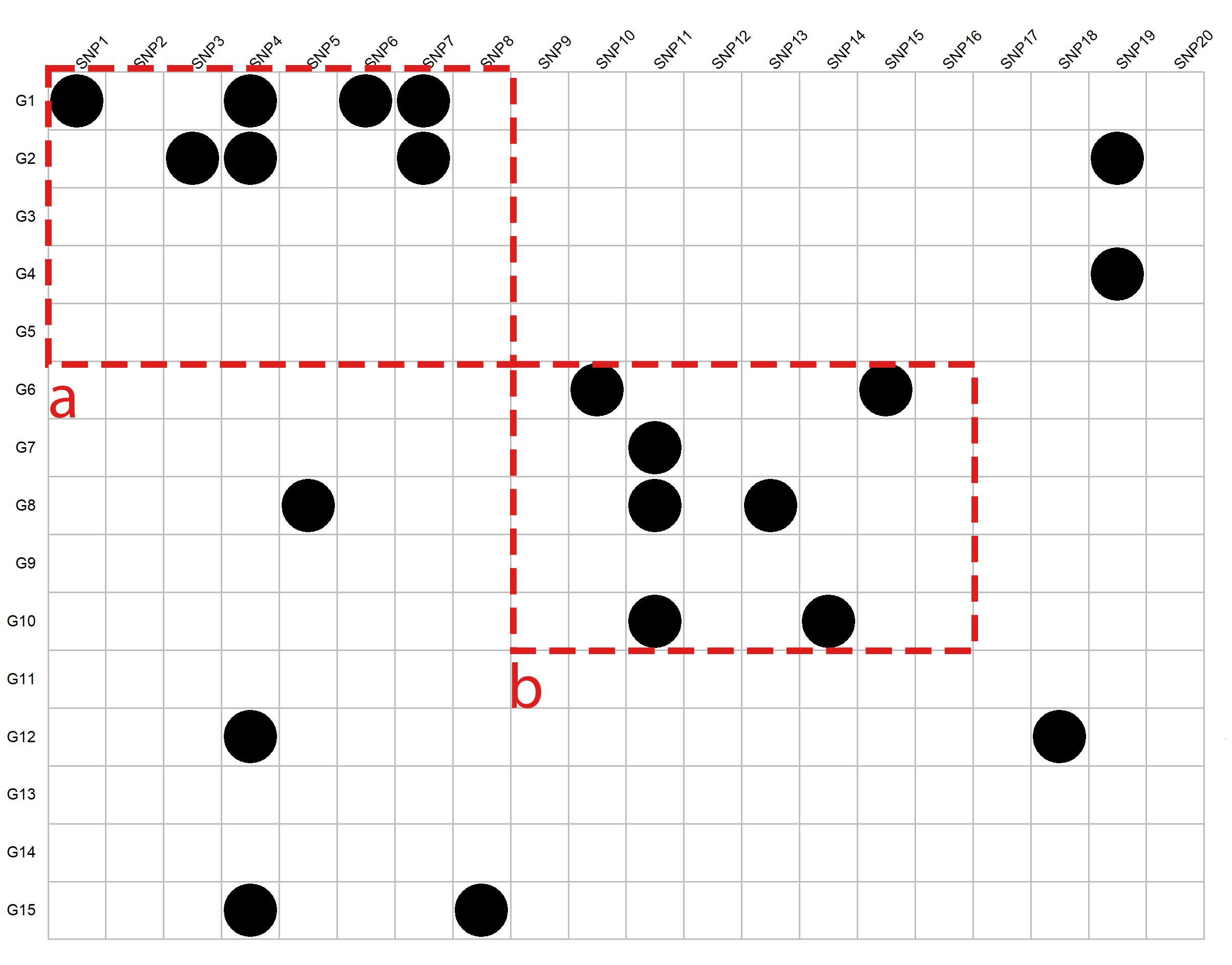}
	\caption{\footnotesize{Toy example of a small dataset with 15 genes and 20 SNPs, truly correlated probes are marked with a black circle. The two-way feature sets are indicated by red rectangles have the same pair-TDP of 7/40. However, number of rows including at least one discovery is different for the two sets.}}
	\label{paiGE TDPtoy}
\end{figure}

While the pair-TDP is a useful overall estimate, it does not specify how many features from omic A are associated with the set of selected features from omic B or the other way around. We can illustrate this with a toy example in Figure \ref{paiGE TDPtoy}. This shows a $15 \times 20$ matrix of pairwise associations. Two two-way feature sets are marked in the matrix, both of size $5\times 8$, denoted by $a$ and $b$. For both sets the pair-TDP is 7/40, yet it is very clear that the structure of associations in these sets is very different. For set $a$, all true associations are concentrated in two of the genes, while for set $b$ the associations are more scattered, indicating that more of the features in omic A are associated with the feature set of omic B. 

In this section, therefore, we will introduce two new TDP metrics which will be able to quantify the signal in the two-way feature sets that focus specifically on both contributing omics. Since the two omics form the rows and columns of the matrix $M$ we will refer to these as row-TDP and column-TDP. Due to symmetry it suffices to discuss row-TDP only. 

In words, the row-TDP of a two-way feature set $S_A \times S_B$ is the proportion of features from $S_A$ that associates with at least one feature in $S_B$. In Figure \ref{paiGE TDPtoy}, row-TDP is 2/5 for two-way feature set $a$ and 4/5 for two-way feature set $b$. Column-TDP is 5/8 both for $a$ and $b$. By reporting row-TDP and column-TDP in addition to pair-TDP, the researcher gets a much better impression of the structure of the associations. Both row-TDP and column-TDP are always at least as large as the pair-TDP.

Formally, for $j = 1, \ldots, p$, let $V_j = \{j\} \times \{1,\ldots,q\}$ be the $j$th row of $M$. We can define row-TDP in terms of set $\times $ set null hypotheses as follows. We may see the two-way feature set $S = S_A \times S_B$ as composed of rows. There are $|S_A|$ such rows, defined as $S_{\mathrm{row}(j)} = S \cap V_j = \{j\} \times S_B$ for each $j \in S_A$. Such row contains signal if $d(S_{\mathrm{row}(j)})>0$; we call these the true discoveries at the row level. The number of rows that contain signal is
\begin{equation} \label{drdef}
d^r(S) = | j \in S_A\colon d(V_j)>0|,
\end{equation}
and the row-TDP of $S$ is therefore $d^r(S)/|S_A|$. 

A lower bound for $d^r(S)$ can be derived using the closed testing-based arguments used by \cite{Goeman2011}, as
\begin{equation} \label{eq row TDP}
\bar d^r(S) = |S_A| - \max_{I \subseteq S_A} \{|I|\colon 
\bar d(I \times S_B)=0 \}.
\end{equation}
To see that this lower bound is valid, let $E = \{\bar d(S) \leq d(S) \textrm{\ for all $S$}\}$ be the event that SEA did not make an error. By (\ref{simbounds}), we have $\mathrm{P}(E) \geq 1-\alpha$. If $E$ happened, then we have
\begin{eqnarray*}
\bar d^r(S) &\leq& |S_A| - \max_{J \subseteq S_A} \{|J|\colon 
d(J \times S_B)=0 \} \\
&=& |S_A| - | j \in S_A\colon d(V_j)=0|\\
&=& d^r(S).
\end{eqnarray*}
Therefore, 
\[
\mathrm{P}(\bar d^r(S) \leq d^r(S)) \geq \mathrm{P}(E) \geq 1-\alpha,
\]
which implies that $\bar d^r(S)$ is indeed a valid lower bound to the row-TDP of $S$. Moreover, we even have simultaneous control:
\begin{equation} \label{row TDP property}
\mathrm{P}(\bar d^r(S) \leq d^r(S) \textrm{\ for all $S$}) \geq \mathrm{P}(E) \geq 1-\alpha,
\end{equation}
since the coverage event $E$ is the same for all $S$. 

We can bound the column-TDP in the same way. Since the coverage event $E$ is the same for row-TDP and for column-TDP, we get simultaneous control for row-TDP and column-TDP, implying that we do not need to correct for multiplicity if we want to report both. In fact, the coverage event is also the same for pair-TDP, so all three can always safely be reported together.

\section{Algorithms}\label{SSalg}

In this section we introduce an algorithm targeted to finding the quantity $\bar{d^r}(S)$. The algorithm consists of two steps. The first step, which we call the single step shortcut, brackets $\bar{d^r}(S)$. The second step uses a branch-and-bound algorithm that narrows these brackets further and further until convergence.

To calculate $\bar{d^r}(S)$, according to (\ref{eq row TDP}), we need to check for many sets $T$ whether $\bar{d^r}(T)>0$ in SEA. \cite{Goeman2019} showed that the null hypothesis $\bar d(T)>0$ if and only if for some $1\leq r \leq |T|$, we have
	\begin{equation}\label{simestest}
	 hp_{(r:T)}\leq r\alpha,
	\end{equation}
where $p_{(r:T)}$ is the $r$th ordered $p$-value with index in $T$ and $h$ is defined in (\ref{functionh}). The calculations in (\ref{simestest}) essentially use discretized $p$-values, since it is only relevant how many multiples of $\alpha/h$ each $p$-value makes. Defining the \textit{$p$-category} of each $p$ as $r=\min\{r :h p \le r \alpha \}$, we can equivalently say that $\bar d(T)>0$ if and only if there exists some $u \in \mathbb{N}$ such that there are $u$ or more $p$-categories in $T$ that are at most $u$. We will work with this latter formulation of (\ref{simestest}).

We will illustrate our algorithm with a toy example of a $6 \times 7$ two-way set. It is presented in terms of $p$-categories in Table \ref{ex1}. Here, for example the top right $p$-category of 3 indicates that the corresponding $p$-value is between $2\alpha/h$ and $3\alpha/h$. We can see that $\bar d(V_1)>0$, since there is a $p$-category of 1 in the first row, and that $\bar d(V_4)>0$, since there are two $p$-category of 2 in the fourth row. All other rows have $\bar d(V_j)=0$.

\begin{table}[ht]
\centering
    \begin{tabular}{clccccccl}
        \hline
        \textbf{$V_1$} &  & 3  & 948 & 35 & 5 & 14 & 1 & 24  \\
        \textbf{$V_2$} &  & 11 & 49 & 7 & 2  & 27 & 224 & 18   \\
        \textbf{$V_3$} &  & 13  & 160 & 20 & 12 & 4 & 2 & 8   \\
        \textbf{$V_4$} &  & 78  & 2 & 75 & 3 & 5 & 25 & 2 \\
        \textbf{$V_5$} &  & 17  & 4 & 142 & 80 & 15 & 451 & 31   \\
        \textbf{$V_6$} &  & 82  & 71 & 23 & 67 & 762 & 5 & 20   \\     
        \hline
    \end{tabular}
\caption{Toy example - $p$-categories matrix}
\label{ex1}
\end{table}

To see this more easily, we can represent the same table in terms of cumulative categories, as shown in Table \ref{ex2}. Per row, the cumulative categories $\{c_{jk}\le u ; u \in \mathbb{N} \}$ count the number of $p$-categories below $u=1,2,\ldots$. In this table, we can read off more directly which rows have $\bar d(V_j)>0$, since these rows have $c_{jk}\geq k$ for some $k$. For example, we see that $\bar d(V_1)>0$ since $c_{11}\geq 1$, and that $\bar d(V_4)>0$ since $c_{42}\geq 2$. The cumulative categories table never needs more than $|S|$ columns, and is further bounded by \citet[Lemma 3]{Goeman2019}.

\begin{table}[ht]
\centering
\begin{tabular}{clccccccll}
\multicolumn{1}{l}{} & \multicolumn{1}{c}{} & \multicolumn{8}{c}{u} \\ \cline{3-10} 
\multicolumn{1}{l}{\textbf{}} & \textbf{} & \multicolumn{1}{l}{\textbf{1}} & \multicolumn{1}{l}{\textbf{2}} & \multicolumn{1}{l}{\textbf{3}} & \multicolumn{1}{l}{\textbf{4}} & \multicolumn{1}{l}{\textbf{5}} & \multicolumn{1}{l}{\textbf{6}} & \textbf{7} & ... \\ \hline
\textbf{$c_{1k}$} &  & 1 & 1 & 2 & 2 & 3 & 3 & 3  & ... \\
\textbf{$c_{2k}$} &  & 0 & 1 & 1 & 1 & 1 & 1 & 2  & ... \\
\textbf{$c_{3k}$} &  & 0 & 1 & 1 & 2 & 2 & 2 & 2  & ... \\
\textbf{$c_{4k}$} &  & 0 & 2 & 3 & 3 & 4 & 4 & 4  & ... \\
\textbf{$c_{5k}$} &  & 0 & 0 & 0 & 1 & 1 & 1 & 1  & ...  \\
\textbf{$c_{6k}$} &  & 0 & 0 & 0 & 0 & 1 & 1 & 1  & ...  \\
\hline
\end{tabular}
\caption{Toy example - Cumulative $p$-category count matrix}
\label{ex2}
\end{table}

We will now use this cumulative table to construct both an upper and a lower bound for $\bar d^r(S)$. We will find $B$ and $H$ such that $B \leq \bar d^r(S) \leq H$.

According to (\ref{eq row TDP}), to calculate $\bar d^r(S)$, it is not sufficient to check whether $\bar d(V_j)>0$ for all $j$, but we also need to look at unions, e.g.\ $\bar d(V_j \cup V_t)$. From Table \ref{ex2} we can easily check whether $\bar d$ is positive for such unions by adding the corresponding rows. For example, though $\bar d(V_2)=0$ and $\bar d(V_3)=0$, we see that $\bar d(V_2 \cup V_3)>0$, since $c_{22}+c_{32} =2$. For every $J \subseteq S_A$, define $c_{J,k} = \sum_{j\in J} c_{jk}.$

To find the bound $B$, we sort the columns of the cumulative categories table in ascending order and take a cumulative sum in each column. In the toy example, the result of this operation is given in Table \ref{ex3}. Call the elements of the resulting table $w_{jk}$. Now, we have the property that, if $|J|=j$, then $c_{J,k} \geq w_{jk}$. To see that this is true, note that each $w_{jk}$ is the sum of the smallest elements in the corresponding column of the cumulative categories table. It follows that, if $w_{jk} \geq k$ for some $k$, then $c_{J,k} \geq k$ for all $J$ with $|J|=j$, so $\bar d(J \times S_B) >0$. In fact, the same automatically holds for all $J$ with $|J|\geq j$, since $w_{jk}$ is increasing in $j$. Therefore, the maximal $|J|$ for which $\bar d(J \times S_B) =0$ is at most $j-1$, so, by (\ref{eq row TDP}), $\bar d^r(S) \geq |S_A| - (j-1)$. To exploit this fact in an optimal way, we must look for the smallest $j$ such that $w_{jk} \geq k$. In Table \ref{ex3}, we see that $w_{3k} < k$ for all $k$, but $w_{42} = 2$, $w_{44}=4$ and $w_{45}=5$, so $j=4$ and $\bar d^r(S) \geq |S_A| - (4-1) = 3$.

\begin{table}[ht]
\centering
\renewcommand{\arraystretch}{1.4}
\begin{tabular}{l l r *{11}{c}}
\multicolumn{1}{l}{} & & \multicolumn{7}{c}{u}  \\ \cline{3-10} 
\multicolumn{1}{l}{\textbf{}} & \textbf{} & \multicolumn{1}{c}{\textbf{1}} & \multicolumn{1}{c}{\textbf{2}} & \multicolumn{1}{c}{\textbf{3}} & \multicolumn{1}{c}{\textbf{4}} & \multicolumn{1}{c}{\textbf{5}} & \multicolumn{1}{l}{\textbf{6}} & \textbf{7} & ... \\ \hline
\textbf{$w_{1k}$} & & 0 & 0 & 0 & 0 & 1 & 1 & 1 & ...\\
\textbf{$w_{2k}$} & & 0 & 0 & 0 & 1 & 2 & 2 & 2 & ...\\
\textbf{$w_{3k}$} & & 0 & 1 & 1 & 2 & 3 & 3 & 4 & ...\\
\textbf{$w_{4k}$} & & 0 & \circled{2} & 2 & \circled{4} & \circled{5} & 5 & 6 & ...\\
\textbf{$w_{5k}$} & & 0 & 3 & 4 & 6 & 8 & 8 & 9 & ...\\
\textbf{$w_{6k}$} & & 1 & 5 & 7 & 9 & 12 & 12 & 13 & ...\\
\hline

\end{tabular}
\caption{Toy example - Calculation of bound $B$. The smallest $j$ for which $w_{jk} \geq k$, for some $k$, is 4.}
\label{ex3}
\end{table}

To obtain the heuristic $H$, we reorder the rows of the cumulative category table in any way we like, and take the cumulative sum as above. Call the values in the resulting table $v_{jk}$. Each $v_{jk}$ is equal to $c_{J,k}$ for a specific $J$ with $|J|=j$. If, for some $j$, $v_{jk} < k$ for all $k$, then there exists an $J \subset S_A$ such that $\bar d(J\times S_B) = 0$, so, by (\ref{eq row TDP}), $\bar d^r(S) \leq |S_A|-j$. It follows that we need to find the largest $j$ such that $v_{jk} < k$ for all $k$. Since $v_{jk}$ is increasing in $k$, we have $j=j^*-1$, where $j^*$ is the smallest $v_{jk}$ such that $v_{jk} \geq k$ for some $k$. Then $\bar d^r(S) \leq |S_A|-(j^*-1)$. In Table \ref{ex4}, we see that $v_{1k} < k$ for all $k$, but $v_{21} = 1$, so $j=2$ and $\bar d^r(S) \leq |S_A| - (2-1) = 5$.

\begin{table}[ht]
\centering
\renewcommand{\arraystretch}{1.4}
\begin{tabular}{l l c *{11}{c}}
\multicolumn{1}{c}{} & & \multicolumn{7}{c}{u} \\ \cline{3-10} 
\multicolumn{1}{c}{\textbf{}} & \textbf{} & \multicolumn{1}{c}{\textbf{1}} & \multicolumn{1}{c}{\textbf{2}} & \multicolumn{1}{c}{\textbf{3}} & \multicolumn{1}{c}{\textbf{4}} & \multicolumn{1}{c}{\textbf{5}} & \multicolumn{1}{c}{\textbf{6}} & \textbf{7} & ...\\ \hline
\textbf{$v_{1k}$} & & 0 & 0 & 0 & 1 & 1 & 1 & 1 & ...\\
\textbf{$v_{2k}$} & & \circled{1} & 1 & 2 & 3 & 4 & 4 & 4 & ...\\
\textbf{$v_{3k}$} & & 1 & 1 & 2 & 3 & 5 & 5 & 5 & ...\\
\textbf{$v_{4k}$} & & 1 & 3 & 5 & 6 & 9 & 9 & 9 & ...\\
\textbf{$v_{5k}$} & & 1 & 4 & 6 & 7 & 10 & 10 & 11 & ...\\ 
\textbf{$v_{6k}$} & & 1 & 5 & 7 & 9 & 12 & 12 & 13 & ...\\ \hline
\end{tabular}
\caption{Toy example - Calculation of heuristic $H$. The smallest $j$ for which $v_{jk} \ge k$, for some $k$, is 2.}
\label{ex4}
\end{table}

Combining the bound and the heuristic, we find that $3 \leq \bar d^r(S) \leq 4$. Note the similarity between the calculation of $H$ and $B$. Both bounds result from the same algorithm applied to a differently pre-processed table. This algorithm can be performed in linear time in the number of rows and columns of the table, by starting at bottom left, going up in the table when $v_{jk}\geq k$ (or $w_{jk} \geq k$, respectively) and right otherwise. The row $j$ in which we drop of the right edge of the table is the largest $j$ for which $v_{jk} < k$ for all $k$.

The upper bound $H$ is valid for any ordering of the rows, but it can be tighter or less tight depending on the chosen order. To get a small $H$, we need to have rows with small categories concentrated at bottom. To achieve this, we suggest to order the rows by decreasing adjusted $p$-values of the $V_j$'s. Calculation of adjusted $p$-values is given in \citet[equation 5]{Ebrahimpoor2019}.

The search for $H$ and $B$ is summarized in Algorithm \ref{SS}. 

\begin{algorithm}
	\caption{}\label{SS}
	\begin{algorithmic}[0]
    \State \textbf{Input}: Matrix of cumulative $p$-categories with elements $c_{jk}$ and size $r \times l$
    \texttt{\\}
     \Function{\textit{Findj}}{$t_{jk}$}
        \State $j=r$
        \State $k=1$
        \While{$k \leq l$ \& $j \geq 1$}
            \If{$t_{jk} \geq k$}
        \State $j=j-1$
            \Else
        \State $k=k+1$
          \EndIf
        \EndWhile
        \State \Return $\textit{j}$
    \EndFunction
    \texttt{\\}
        \Function{\textit{bound}}{$c_{jk}$} \Comment{Find B}
        \State \hspace*{\algorithmicindent} $w_{jk}  \gets$ Sort each column of $c_{jk}$ in ascending order 
        \State \hspace*{\algorithmicindent} $w_{jk} \gets \sum_{j=1}^{j} w_{jk}$ 
        \State \hspace*{\algorithmicindent} $B=l-\textit{Findj}(w_{jk})+1$
        \State \Return $B$
    \EndFunction
    \texttt{\\}  
        \Function{\textit{heuristic}}{$c_{jk}$} \Comment{Find H}
        \State \hspace*{\algorithmicindent} $v_{jk}  \gets$ Sort rows of $c_{jk}$ by $p^{j}_{adj}$ in descending order 
        \State \hspace*{\algorithmicindent} $v_{jk} \gets \sum_{j=1}^{j} v_{jk}$ 
        \State \hspace*{\algorithmicindent} $H=l-\textit{Findj}(v_{jk})+1$
        \State \Return $H$
    \EndFunction
    \texttt{\\}
        \State \Return $[B, H]$
	\end{algorithmic}
\end{algorithm}

When using the result of the algorithm, the lower bound $B$ is of primary interest, since replacing $\bar d^r(S)$ by the lower bound retains the crucial property \eqref{row TDP property}. The upper bound tells us how well we approximated $\bar d^r(S)$. The precision can be increased by spending more computational power, as we will show in the next section.

\subsection{Branch and bound} \label{unsure}

If $B=H$, the algorithm gives an exact result for $\bar d^r(S)$. If $B<H$, $\bar d^r(S)$ is only bracketed. The brackets can be narrowed by a branch-and-bound algorithm. We present this algorithm in this section.

Branch and Bound is an optimization technique \citep{Land1960, Mitten1970} for discrete problems. It divides the search space into smaller disjoint sub-spaces (branches), evaluates them using a heuristic, and prunes branches that cannot lead to better solutions by using a bound. By repeatedly branching and eliminating unpromising paths, it efficiently searches for the optimal solution. 

Branch-and-bound is useful in our problem, because \eqref{eq row TDP} represents a discrete optimization problem in the exponential search space of all subsets of $S_A$. We repeatedly split the space by either considering subsets including a certain row, or subsets excluding that row. The row to exclude is chosen is the one with the smallest adjusted $p$-value. We apply Algorithm \ref{SSalg} in each subspace evaluated by the algorithm. Here, $H$ plays the role of the heuristic, and $B$ of the bound. 

\begin{table}[ht]
    \centering
    \renewcommand{\arraystretch}{1.2}
    \begin{tabular}{lcccccccllccccccc}
		& \multicolumn{7}{c}{u} &  &  & \multicolumn{7}{c}{u}
		\\ \cline{2-8} \cline{11-17} 
		& 1 & 2 & 3 & 4 & 5 & 6 & 7  &  & & 1 & 2 & 3 & 4 & 5 & 6 & 7 \\
		\cline{1-8} \cline{10-17} 
		$w_{1k}^{+1}$ & \circled{1} & 1 & 2 & 2 & 3 & 3 & 3 & & $v_{1k}^{+1}$ & \circled{1} & 1 & 2 & 2 & 3 & 3 & 3  \\
		$w_{2k}^{+1}$ & 1 & 1 & 2 & 2 & 4 & 4 & 4 & & $v_{2k}^{+1}$ & 1 & 2 & 3 & 4 & 5 & 5 & 5  \\
		$w_{3k}^{+1}$ & 1 & 1 & 2 & 3 & 5 & 5 & 5 & & $v_{3k}^{+1}$ & 1 & 3 & 4 & 5 & 6 & 6 & 7  \\
		$w_{4k}^{+1}$ & 1 & 2 & 3 & 4 & 6 & 6 & 7 & & $v_{4k}^{+1}$ & 1 & 5 & 7 & 8 & 10 & 10 & 11  \\
		$w_{5k}^{+1}$ & 1 & 3 & 4 & 6 & 8 & 8 & 9 & & $v_{5k}^{+1}$ & 1 & 5 & 7 & 9 & 11 & 11 & 12\\
		$w_{6k}^{+1}$ & 1 & 5 & 7 & 9 & 12 & 12 & 13 & & $v_{6k}^{+1}$ & 1 & 5 & 7 & 9 & 12 & 12 & 13\\
		\cline{1-8} \cline{10-17}
	\end{tabular}
	\caption{Toy example - Calculation of H and B for the branch where the first row from $c_{ij}$ matrix is forced to be included in all combinations (set to $w_{1k}^{+1}$ and $v_{1k}^{+1}$). This branch must be discarded based on the value of its bound $B=6$; meaning it is not possible to find an improved Heuristic in this branch.}
\label{exbb1}
\end{table}

\begin{table}[ht]
\centering
    \renewcommand{\arraystretch}{1.2}
    \begin{tabular}{lcccccccllccccccc}
    & \multicolumn{7}{c}{u} &  &  & \multicolumn{7}{c}{u}
    \\ \cline{2-8} \cline{11-17} 
 & 1 & 2 & 3 & 4 & 5 & 6 & 7  &  & & 1 & 2 & 3 & 4 & 5 & 6 & 7 \\
 \cline{1-8} \cline{10-17} 
 $w_{1k}^{-1}$ & 0 & 0 & 0 & 0 & 1 & 1 & 1 &  & $v_{1k}^{-1}$ & 0 & 0 & 0 & 1 
 & 1 & 1 & 1  \\
 $w_{2k}^{-1}$ & 0 & 0 & 0 & 1 & 2 & 2 & 2 &  & $v_{2k}^{-1}$ & 0 & 0 & 0 & 1 
 & 2 & 2 & 2  \\
 $w_{3k}^{-1}$ & 0 & 1 & 1 & 2 & 3 & 3 & 4 &  & $v_{3k}^{-1}$ & 0 & 1 & 1 & 2 
 & 3 & 3 & 4  \\
 $w_{4k}^{-1}$ & 0 & \circled{2} & 2 & 4 & \circled{5} & 5 & 6 &  & $v_{4k}^{-1}$ & 0 & 
 \circled{2} & 2 & 4 & \circled{5} & 5 & 6  \\
 $w_{5k}^{-1}$ & 0 & 4 & 5 & 7 & 9 & 9 & 10 &  & $v_{5k}^{-1}$ & 0 & 4 & 5 & 7 
 & 9 & 9 & 10\\
 \cline{1-8} \cline{10-17}
\end{tabular}
\caption{Toy example - Calculation of $H$ and $B$ for the branch where first row from $c_{jk}$ matrix is removed. As $v_{42}^{-1} \ge 2$ and $v_{45}^{-1} \ge 5$ the improved heuristic is $H=3$  leading to a sure outcome of $3$.}
\label{exbb2}
\end{table}

Going back to the toy example, we can see that it is possible to improve the resulting bound by branching based on first row. Table \ref{exbb1} shows the search for $B$ and $H$ in the subspace where first row is forced to be included; which is achieved by fixing it on the top row throughout the ordering steps of the algorithm. Considering the value of $B=6$, it is not possible to find a better Heuristic in this subspace. Table \ref{exbb2} shows the search for $B$ and $H$ in the subspace where first row is removed. The new Heuristic is smaller than the one calculated based on the full set and leads to an exact outcome of $d^r(S)=3$. 

The resulting branch-and-bound algorithm is given in Algorithm \ref{SSB}.

\begin{algorithm}
	\caption{ }\label{SSB}
    \begin{algorithmic}[0]
	\State \textbf{Initialize}
	\State \hspace*{\algorithmicindent}		$Q= S$
	\State \hspace*{\algorithmicindent} 	$i= 0$
	\State \hspace*{\algorithmicindent} Get $[bound(S), heuristic(S)]$ from Algorithm \ref{SS}
        \State \hspace*{\algorithmicindent} H=heuristic(S)
\texttt{\\}
	\While{$Q \neq \emptyset $  \& $i<maxi$}
	\State i $\gets$ i+1
	\State $C_i =$ last in the queue \& remove $C_i$ from queue
	\State $B^\prime=max(bound(C_i),bound(C_i^p)^1)$ 
	\If {$H \leq B^\prime$}
	   \State \textit{Do nothing}
	   \Else
	   \State $H= min(H, heuristic(C_i))$ 
	   \State Split $C_i$ as
          \State \hspace*{\algorithmicindent}  $C^{+}=\{C_i: \textrm{first row is forced.}\}$
	     \State \hspace*{\algorithmicindent} $C^{-}=\{C_i : \textrm{first row is removed.}\}$
	   \State Add $C^{+}$ and $C^{-}$ to $Q$ keeping record of their parent set $C_i$.
	\EndIf 
	\State
	\EndWhile
	\State \hspace*{\algorithmicindent} $B = min(H, min[bound(C_i); C_i \in Q)])$
	\State \Return $[B , H]$
        \\
        \State \small{1 $C_i^p$, or the parent set, is the set that $C_i$'s are split from as defined above.}
    \end{algorithmic}
\end{algorithm}

Note that the algorithm needs to repeatedly construct cumulative category count matrices. By some bookkeeping, information from earlier calculated matrices can be reused to speed up the steps of the branch-and-bound algorithm. This is implemented within OCEAN R package.

The algorithm can be run until convergence, i.e.\ until the queue is empty. In that case the calculation of $\bar d^r(S)$ is exact. However, this takes exponential computation time in the worst case. However, we may stop the procedure earlier, e.g. after a fixed number of iterations. This may sometimes still yield an exact outcome. If not, the final bounds will often be tighter than the one from Algorithm \ref{SS}. The resulting improved lower bound will still have the crucial property \eqref{row TDP property}.
 
\section{Application} \label{appl}

To illustrate OCEAN's flexibility, we will use it in the joint analysis of mRNA gene expression (GE) and DNA copy number (CN) data from breast cancer (BRCA, 173 samples)  from The Cancer Genome Atlas (TCGA -- data generated by the TCGA Research Network \texttt{https://www.cancer.gov/tcga}). The CN data available is segmented per sample. To be able to analyse the data, we chose $3\times 10^4$ equally spaced loci covering the entire genome and extracted the corresponding segment's value for each location, yielding a vector of $3\times 10^4$ CN values for each of the samples. The GE data was pre-processed as described in \citet{Menezes2009}, which yielded  $74985$ measurements per sample. To put results in context, we also analyse similar data from 125 colorectal cancer samples (CRC), involving  $75085$ GE features after pre-processing. 

We  calculated the Pearson correlation coefficient for all gene expression and copy number probe pairs per cancer type, computing the corresponding $p$-value matrices (GE by CN) of sizes $75085 \times 30000$ for colon cancer and $74985 \times 30000$ for breast cancer. It is very challenging to draw any conclusion about associations between CN and GE based upon all $2\times 10^9$ pairwise tests, due to multiple testing correction. In addition, selected pairs would have to be put back into a biological context before yielding useful information. For example, CN is often interpreted in terms of segments, as CN changes are observed in stretches of DNA. GE, on the other hand, is often interpreted in the context of biological pathways. So it would be useful to group some rows and columns of the resulting matrix of p-values and interpret these groups, effectively using feature sets instead of the original features.

OCEAN can help by enabling us to do exactly that, in a way that preserves the required multiple testing correction level. Here we will define GE
feature sets based on hallmark pathways \citep{Liberzon2015},  and either chromosome arm or chromosome band for CN. For each feature set, GE TDP (row), CN TDP (column) and pairwise TDP (taking both CN and GE into account) were calculated. 

\begin{figure}[ht]
	\captionsetup{width=\linewidth}
    \centering
	\includegraphics[width=\linewidth,keepaspectratio=true]{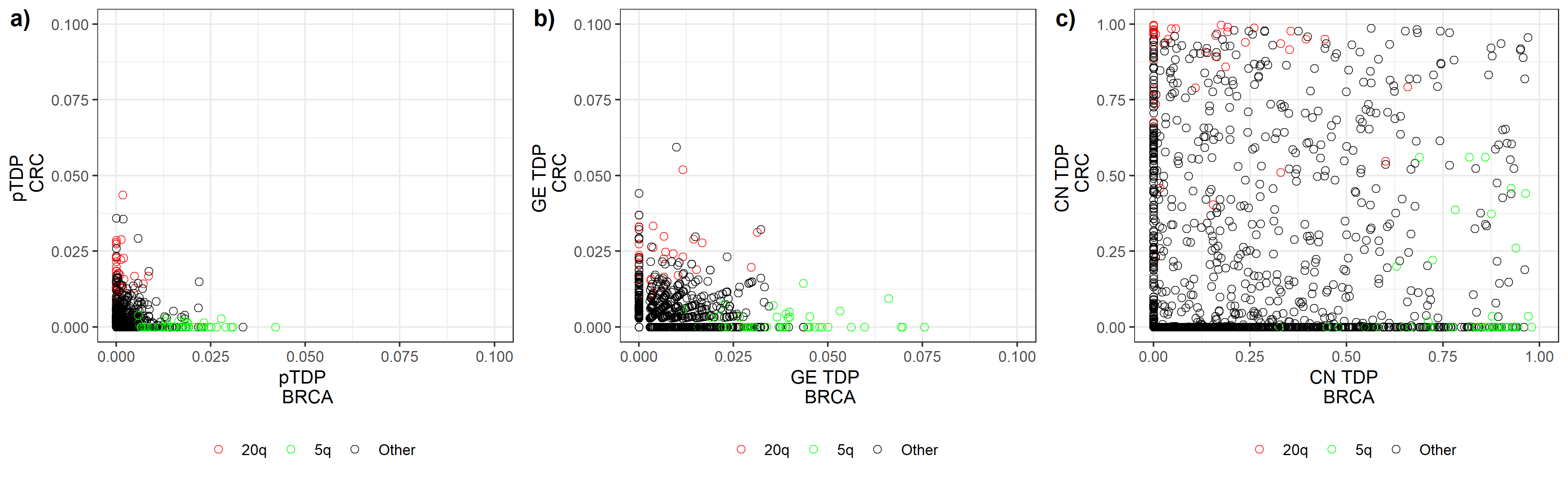}
	\caption{\footnotesize{Scatter plots of TDP for BRCA (x-axis) and CRC (y-axis), for three different TDP types: a) pairwise TDP, b) GE TDP, c) CN TDP. The feature sets used are Hallmark pathways (GE) and chromosome arm (CN). Two chromosome arms (20q and 5q) are marked across the three plots. The graphs are on different scales to make patterns visible.  For the same graphs made using the same scales, see Supplementary Figure 1}}
	\label{fig2}
\end{figure}

Scatter plots in Figure \ref{fig2} compare the results for BRCA and CRC using three TDP types, using hallmark pathways (GE) and chromosome arms (CN) as feature sets. In general, there is consistently more signal from CN (Figure\ref{fig2}c) than from GE (Figure \ref{fig2}b) for both cancer types, with very low true discovery proportions from GE in general, compared to those from CN. The results for pairwise TDP (Figure \ref{fig2}a) do not offer any patterns and seem to reflect mostly the GE signal. 

Some chromosome arms yield clearer patterns. Indeed, feature sets involving 5q  (displayed in green in Figure \ref{fig2}) have relatively high values for all TDPs in BRCA, but for CRC only some of these sets carry a large signal for CN TDP. This is consistent with reports of loss of 5q in samples of a  breast cancer subtype, and this loss includes at least 2 genes involved in a {\sl BRCA1}-dependent DNA repair pathway (see for example \citet{weigman_2012}). As samples presenting this CN change were reported to have poorer prognosis, the CN change is likely to impact GE levels, in particular of genes in Hallmark pathways.

Conversely, feature sets involving 20q (displayed in red in Figure \ref{fig2}) display higher TDPs for CRC  than BRCA. This also makes sense, as DNA copy gain on 20q is often observed for CRC \citep{ried_2019}. This copy number gain is in fact already found in adenomas before progression to carcinoma, suggesting it is likely to impact GE levels. Indeed, it is known that 20q, among other commonly observed CN changes, do have a direct impact on GE levels (see references in \citet{ried_2019}).

\begin{figure}[htbp]
	\captionsetup{width=\linewidth}
    \centering
	\includegraphics[width=\linewidth,keepaspectratio=true]{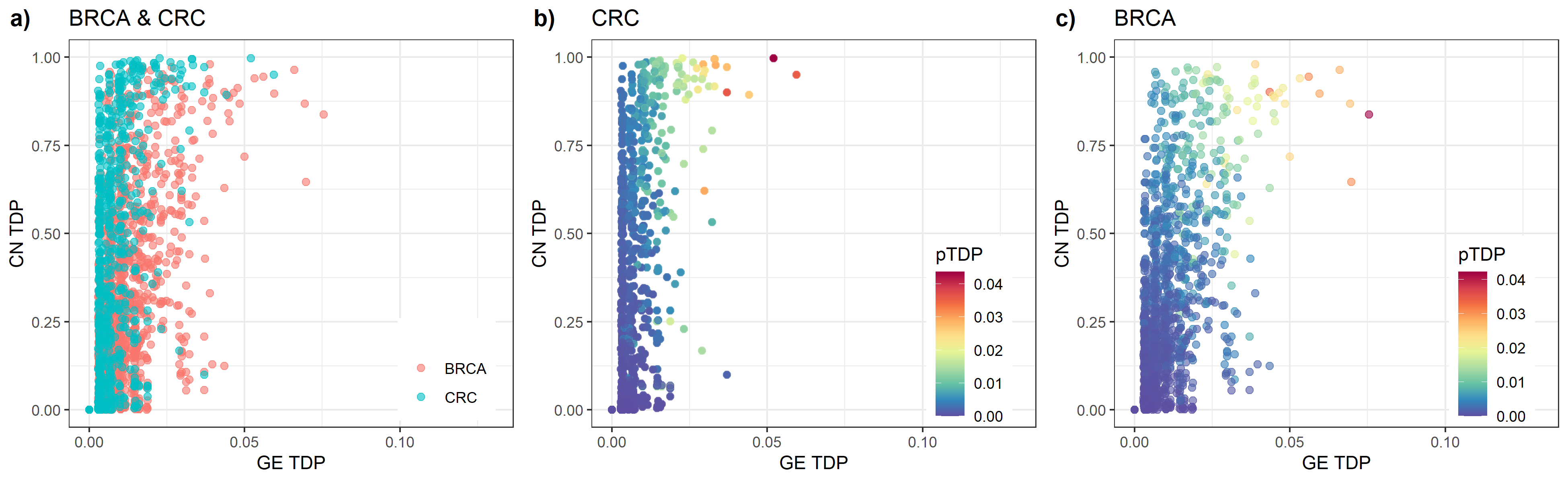}
	\caption{\footnotesize{Scatter plots of CN TDP (y-axis) and GE TDP (x-axis) for CRC and BRCA: a) overlayed results of BRCA and CRC, b) results from CRC only, coloured by their pairwise TDP, c) results from BRCA only, coloured by pairwise TDP. As before, Hallmark pathways (GE) and chromosome arms (CN) are the feature sets used.}}
	\label{fig3}
\end{figure}

As shown in Figure \ref{fig3}, CRC samples yield stronger CN effect (as measured by the pairwise TDP in association with GE) than BRCA. This is in accordance with previous studies where GE and CN have shown stronger associations in CRC than in BRCA \citep{Menezes2016}. Once more, it is evident that pairwise TDP, although correlated with GE TDP and CN TDP, does not offer sufficient specificity in these data. 

\begin{figure}[htbp]
	\captionsetup{width=\linewidth}
    \centering
	\includegraphics[width=\linewidth,keepaspectratio=true]{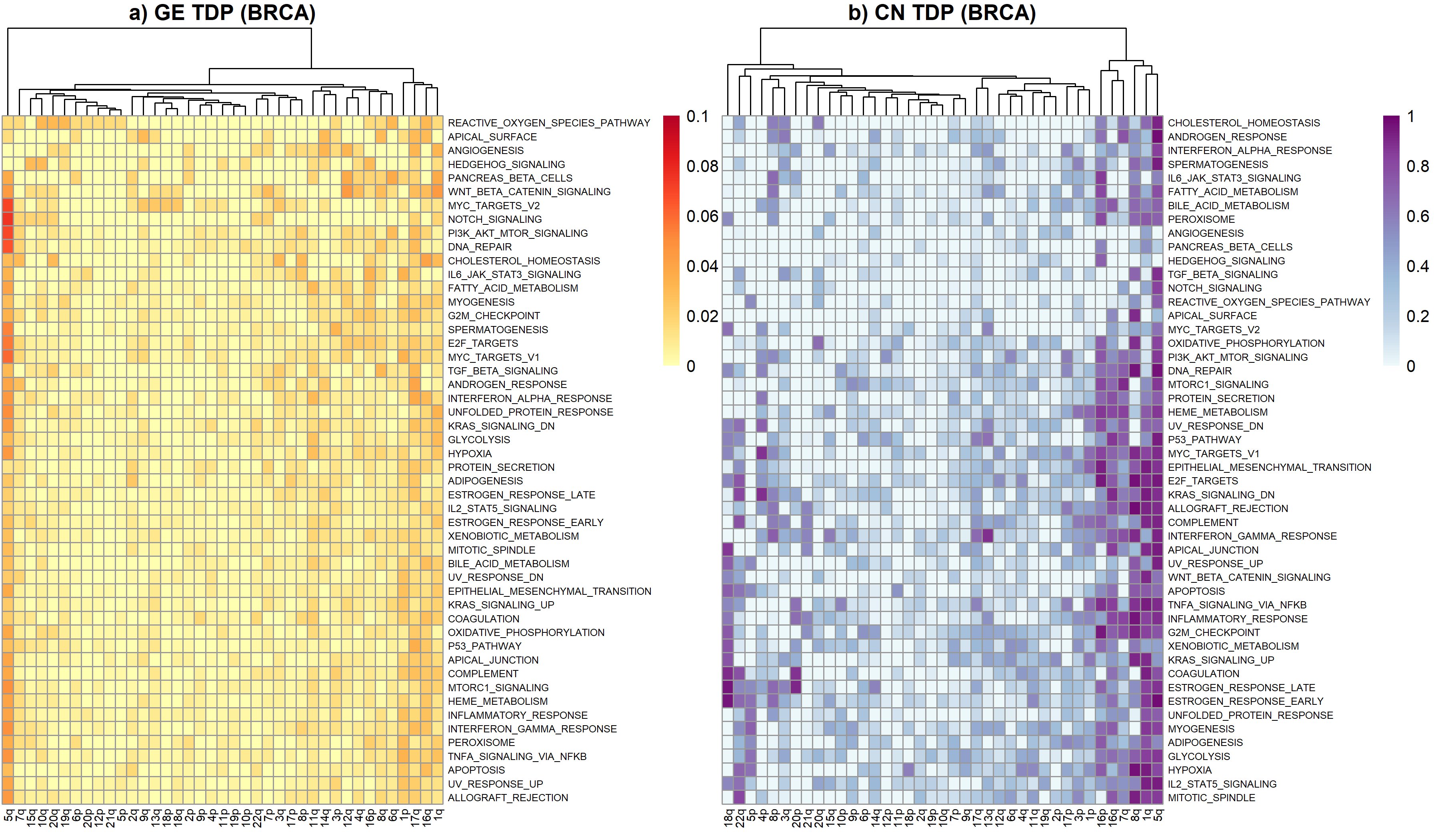}
	\caption{\footnotesize{Hierarchically clustered heatmap of a) GE TDP and b) CN TDP for BRCA. As before, Hallmark pathways (GE) and chromosome arms (CN) are the feature sets used. }}
	\label{fig4}
\end{figure}

Figure \ref{fig4} displays  heatmaps of GE TDP (a) and CN TDP (b) for two-way feature sets defined by Hallmark pathways (rows) and chromosome arms (columns) for the BRCA data. Some patterns are shared between both TDP levels, whilst others are specific to one TDP type. For instance, 5q shows consistently high CN and GE TDPs for many Hallmark pathways (see left-most column in figure \ref{fig4}a and right-most column in figure \ref{fig4}b), whereas 1q and 8q only display high CN TDP with most pathways. This means that a high proportion of CN measurements within 1q and 8q are found to be associated with several Hallmark pathways, even after multiplicity correction, but the same does not happen with GE measurements. Similar plots for CRC (Supplementary Figure 2) show that GE and CN TDPs for 20q have high values for many Hallmark pathways. On the other hand, CN TDP for 18q and 13q is high for most Hallmark pathways, but the same is not true for GE TDPs.

\begin{figure}[htbp]
	\captionsetup{width=\linewidth}
    \centering
	\includegraphics[width=\linewidth,keepaspectratio=true]{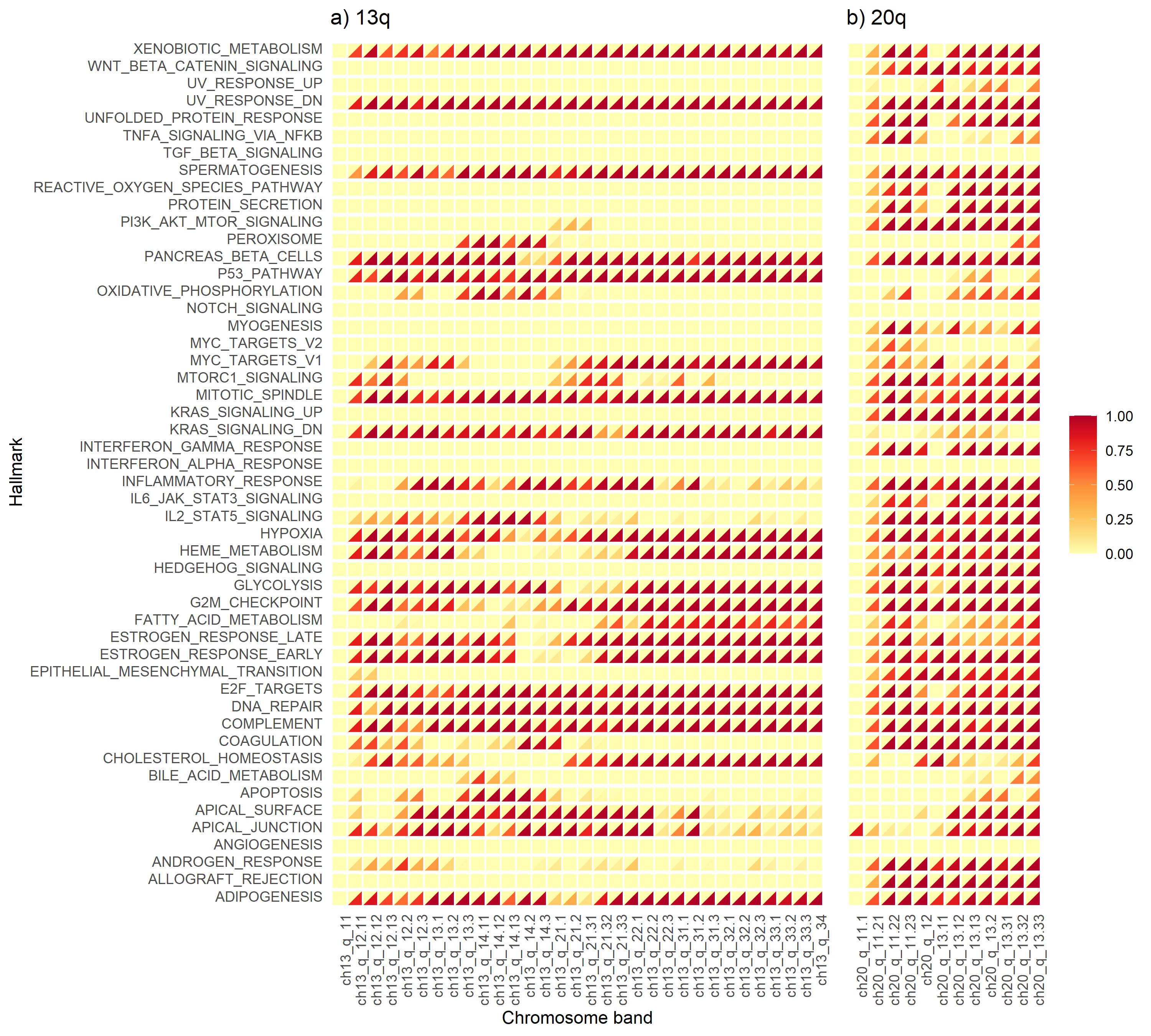}
	\caption{\footnotesize{TDPs for CRC results, GE TDP on the upper triangle and CN TDP on the lower triangle per feature set pair. Hallmark pathways (GE) and chromosome bands (CN) are the feature sets used. The figure provides details at chromosome band level for the two prominent chromosome arm from Supplementary Figure 2: a) 13q and b) 20q.}}
	\label{fig5}
\end{figure}

Chromosome arms are useful CN feature sets, as there are CN aberrations which span an entire arm. However, there are also smaller CN aberrations, which are then likely missed when using an entire chromosome arm. For example, DNA CN gains on 13q and 20q are often observed in CRC \citep{ried_2019}, and it may therefore be useful to further explore results at a finer scale on these chromosome arms. One finer scale is that of chromosome bands, which are smaller regions of DNA. So, we recomputed CN TDPs for the CRC data, now using chromosome bands on 13q and 20q (see \ref{fig5} for CRC, and  Supplementary Figure 3 for the corresponding figure for BRCA using bands in 5q and 1q).  

As before, CN TDPs are typically considerably higher than corresponding GE TDPs, and some Hallmark pathways display consistently high CN TDPs for most chromosome bands. The uniformity of results across bands suggests there is not much heterogeneity between bands, probably because many CN changes span almost the entire chromosome arm. However, there are pathways with CN TDPs high for only regions in 20q, such as \verb+MYC_TARGETS_V1+ (Figure \ref{fig5}). So the use of chromosomal bands helps us zoom into regions within a chromosome arm. A similar plot in the supplementary provides this level of details for 5q and 1q from BRCA (Supplementary Figure 3).

These illustrations highlight the flexibility that OCEAN offers for the analysis of multi-omics data, capturing interesting patterns at different levels of detail, while strictly preserving type I error. Note that we have analyzed 2 different datasets and the family-wise error rate (FWER) control holds for all the results from each of these datasets. Reanalyses, such as using new sets as illustrated above, are quickly performed, as they require merely a recalculation of TDPs.

\section{Implementation}

Despite the seemingly complex nature of the OCEAN algorithm, it is remarkably straightforward to analyze multi-omics data using OCEAN R package available on GitHub (\url{https://github.com/mitra-ep/OCEAN}).

Implementation of OCEAN algorithm involves two main steps: preparing the matrix and calculating discovery rates. The preparation step encompasses calculating the pairwise $p$-value matrix $P$, determining $h$ (as defined in \ref{functionh}), and estimating the number of columns for the cumulative categories matrix $c_{jk}$. These calculations, which utilize the full dimension of the multi-omics data, are the most time-consuming part of the process. However, this step is performed only once for a given omics pair and the results can be used to explore any potential feature sets of interest. For the BRCA dataset, preprocessing step took about 23 minutes.

Once the preparation step is completed, Algorithm \ref{SS} can be adopted to calculate the row-TDP and column-TDP for any given two-way feature set and is quite efficient. It can take between a few seconds to up to eight minutes to calculate row-TDP for a feature set with 100 to over a 1000 rows. Additionally, the pair-TDP can be obtained using the SEA algorithm, alongside the other two discovery rates. The \textit{ocean} function efficiently computes these three TDPs at once for a given feature set. For the CRC dataset, calculation of the TDPs takes around one minute if both number of rows and columns of the feature set are under 600, and it can take up to four minutes for a feature set of size $300 \times 1000$. In case the results are not exact \textit{i.e.} $B \neq H$, \ref{SSB} will be adopted which can take longer to run,depending on the size of feature set and the number of necessary steps for convergence. For instance, for a two-way feature set of size $300 \times 500$, it takes 30 minutes to run the iterative algorithm for 100 steps, and about an hour if the algorithm converges at 500 steps, assuming $B \neq H$ for only row-TDP (GE TDP). However, the rate of such outcomes was only 1.2\% and 4.4\%  for the CRC data based on Hallmark pathways and chromosome arms, and 0.4\% and 0.8\% for BRCA based on Hallmark pathways and chromosome bands, respectively for GE TDP and CN TDP.

\section{Discussion}
OCEAN provides a flexible tool to explore multi-omics datasets in terms of two-way feature sets. While pairwise TDPs, from SEA, effectively summarize associations and control Type I error, interpreting results can be challenging due to the complex association structure inherent to multi-omics data. To provide such insight, in addition to pairwise TDPs, row- and column-TDP metrics were introduced, which separately reflect the correlation pattern for each omics dataset. 

The closed testing framework makes OCEAN flexible in both the number of two-way feature sets being studied, as well as their definitions and eliminates limitation to {\sl a priori} selected feature sets. In particular, it is possible to revise the feature sets used after observing the data, keeping the FWER control level.

This flexibility enhances the utility of OCEAN in exploring associations of genomics features in a data-driven and unbiased manner, accommodating the complexity and dimension of genomics datasets.

OCEAN is not free of assumptions. It assumes positive dependence among features, just like the widely used Benjamini-Hochberg (BH) procedure. OCEAN can therefore safely be used whenever the BH procedure can. 

The example in section \ref{appl} involved studying associations between gene expression and DNA copy number data. In general, OCEAN is not limited to any specific omics data type, and also is not tailored to cancer data. The algorithm can be used to simultaneously analyze any two (or more) omics datasets. Indeed, generalization of OCEAN to more than two dimensions is straightforward in theory, involving three-way feature sets tested in terms of TDPs for each side of a cube. However, extension to more than two feature sets becomes increasingly computationally burdensome due to the high number of associations involved.

We have explored column and row TDPs, within rectangular sets (same number of rows for all columns). However OCEAN can be extended to include  other error types, defined in terms of disjoint subdivisons other than rows or columns. TDP for such other error types can be incorporated within the same closed testing procedure and will not inflate type-I error. 

The proposed approach can be used with any chosen pairwise test to derive the pairwise $p$-value matrix $P$. Therefore, another potential extension is adopting more complex statistical tests, rather than the Pearson's correlation test used in this paper. For example, OCEAN can be used to detect causal or non-linear associations whenever the individual association tests can. 

\section{Acknowledgments}

This research was funded by NWO VIDI grant number 639.072.412. The results presented in section \ref{appl} are based upon data generated by the TCGA Research Network: \url{https://www.cancer.gov/tcga}.

\bibliography{ocean_draft}

\newpage

\appendix

\section{Supplementary Material}

Details of the results are presented in sections 6 of the main article. Here we present some additional figures.\\

\noindent \textbf{Supplementary Figure 1} \\

Scatter plot of TDPs with similar scaling.

\begin{figure}[ht]
	\captionsetup{width=\linewidth}
	\centering
	\includegraphics[width=\linewidth,keepaspectratio=true]{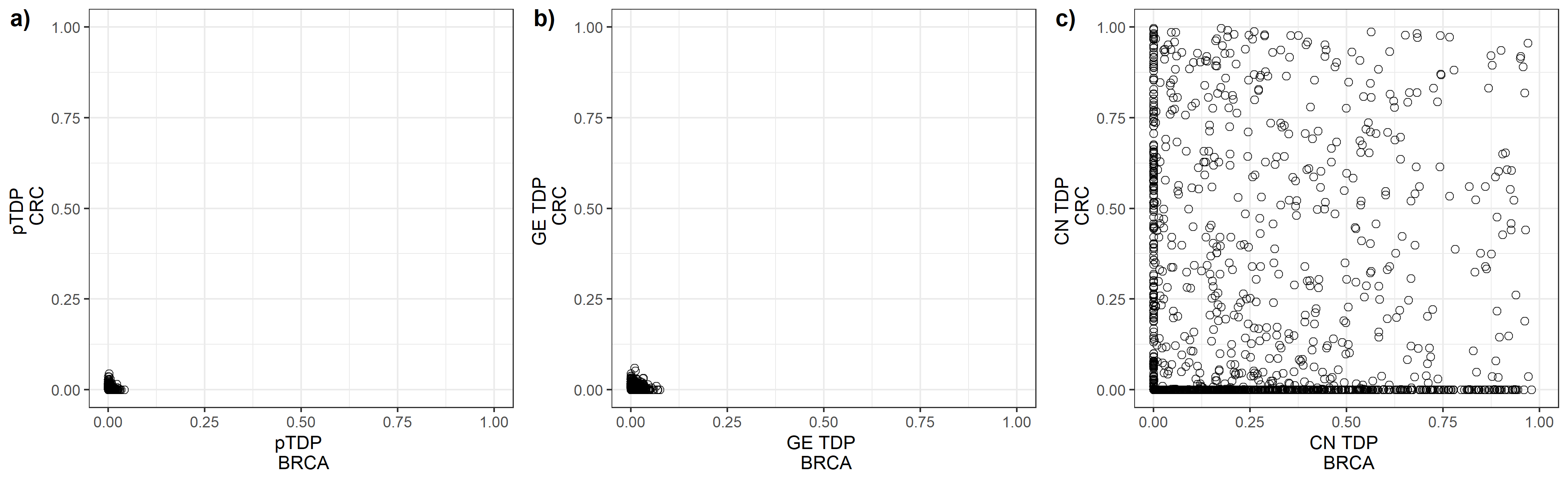}
	\caption{\footnotesize{Scatter plots of TDP at three levels for CRC against BRCA. a) pairwise TDP, b) GE TDP, c) CN TDP. The data are aggregated by hallmark pathways for GE and by chromosome arm for CN.}}
\end{figure}

\noindent \textbf{Supplementary Figure 2} \\

Heatmap of GE TDP aggregated by hallmark pathways (rows) and chromosome arms (columns) for CRC in terms of GE TDP (a) and CN TDP (b).  

\begin{figure}[ht]
	\captionsetup{width=\linewidth}
	\centering
	\includegraphics[width=\linewidth,keepaspectratio=true]{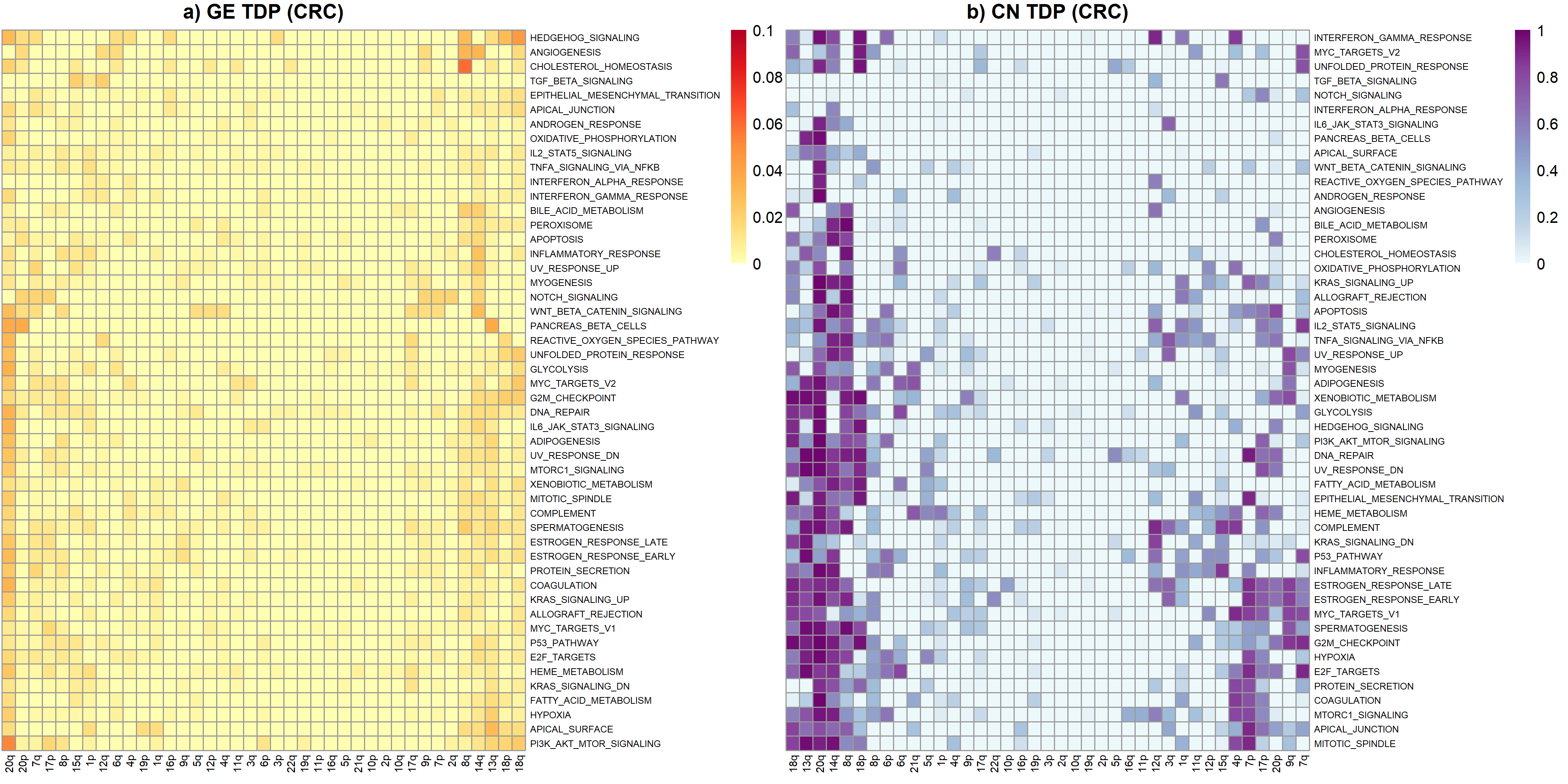}
	\caption{\footnotesize{Hierarchically clustered heatmap of a) GE TDP and b) CN TDP for CRC.}}
\end{figure}

\pagebreak

\noindent \textbf{Supplementary Figure 3} \\

GE TDP and CN TDP for BRCA, with focus on q5 chromosome arm.\\

\begin{figure}[ht]
	\captionsetup{width=\linewidth}
	\centering
	\includegraphics[width=\linewidth,keepaspectratio=true]{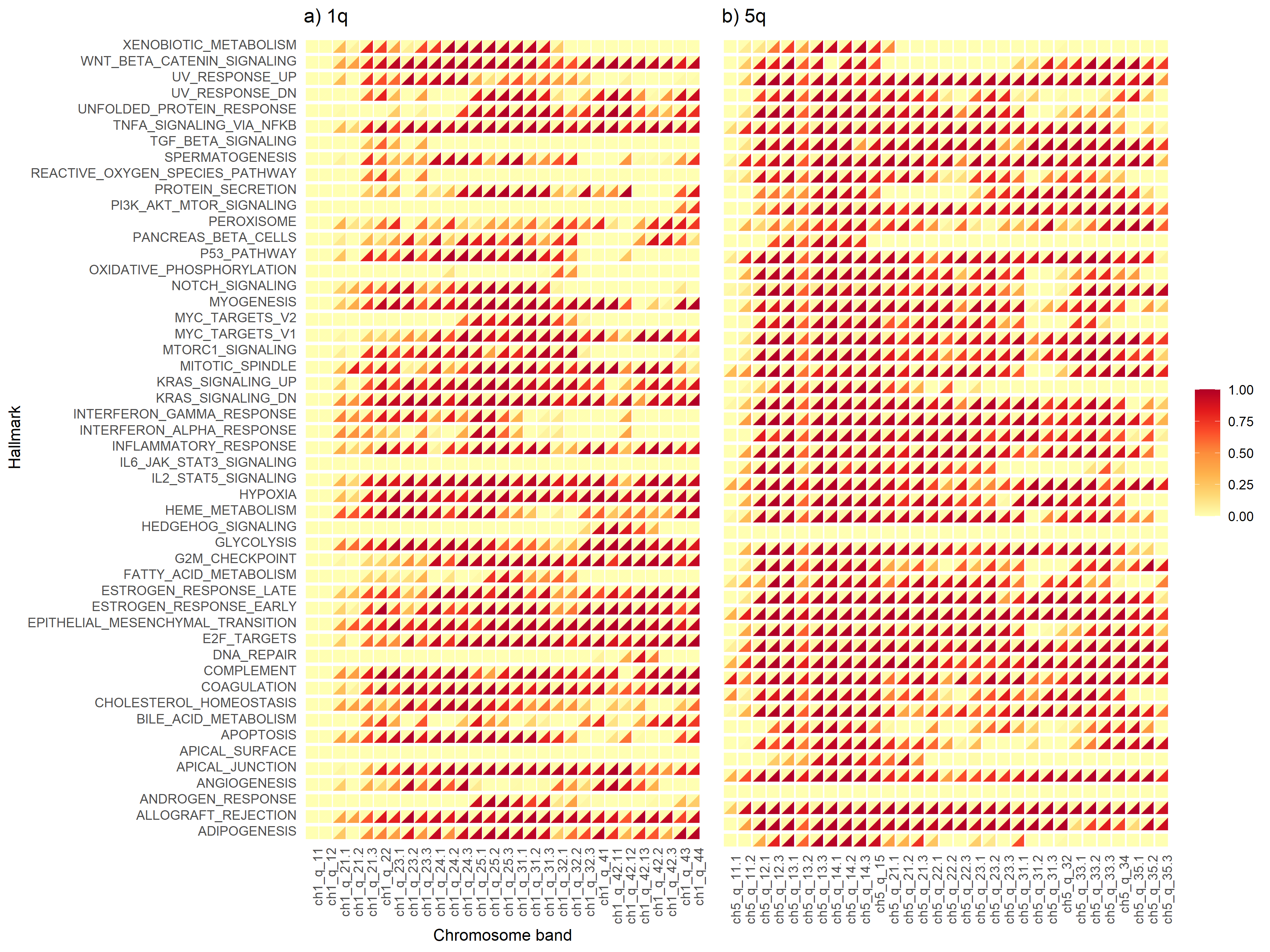}
	\caption{\footnotesize{Correlation matrix of BRCA results with GE TDP on the upper triangle and CN TDP on the lower triangle; CN measurements are aggregated by chromosome band and GE by hallmark pathways. The figure basically zooms into 5q chromosome arm and provides details per chromosome band.}}
\end{figure}

\end{document}